\documentclass[mypaper,8pt,twoside]{CoAst}
\usepackage{epsf,graphicx,fancyhdr,sfmath}
\renewcommand{\chaptermark}[1]{\markboth{#1}{}}

\newcommand{\kopf}{\small\itshape Comm.\ in Asteroseismology, N$^{\textsf{\underline{o}}}$ 159, 2009\\
Proceedings of the JENAM 2008 Symposium N$^{\textsf{\underline{o}}}$~4:
Asteroseismology and Stellar Evolution}

\newcommand{\Authors}[1]{\begin{center}\normalsize\bf\sf #1 \end{center}}

\renewcommand{\author}[1]{\begin{center}\normalsize\bf\sf #1 \end{center}}
\newcommand{\Address}[1]{\begin{center}\small\sf #1 \end{center}}

\renewenvironment{abstract}{\section*{Abstract}\normalsize\sf}{}
\newcommand{\References}[1]{\begin{flushleft}{\large References\\}\vspace*{2mm}\small #1 \end{flushleft}}

\newcommand{\chapterCoAst}[2]{\chapter[\sf\normalsize #1\\ \footnotesize \hspace*{5mm}by #2 \sf\normalsize][]{#1\\}\rhead[\fancyplain{}{\sf\footnotesize \center{#1}}]{\fancyplain{}{\sffamily\thepage}}\lhead[\fancyplain{\kopf}{\sffamily\thepage}]{\fancyplain{\kopf}{\sf\footnotesize \center{#2}}}}

\newcommand{\figureCoAst}[5]{\begin{figure}[#4]
\centering
\includegraphics*[#5]{#1}
\caption{#2}
\label{#3}
\end{figure}}

\renewcommand{\chaptername}{}

\newcommand{\acknowledgments}[1]{\vspace*{5mm}\noindent\textbf{Acknowledgments.} #1}

\def\rfr{\smallskip\par\noindent
       \hangindent=7truemm
       \hangafter=1}

\begin{document}
\sf

\chapterCoAst{Challenges for stellar pulsation and evolution theory}
{Jadwiga Daszy\'nska-Daszkiewicz}
\Authors{Jadwiga Daszy\'nska-Daszkiewicz}
\Address{
Instytut Astronomiczny, Uniwersytet Wroc{\l}awski, Kopernika 11,
 51-622 Wroc{\l}aw, Poland
}

\noindent
\begin{abstract}
During the last few decades, great effort has been made towards
understanding hydrodynamical processes which determine the structure and
evolution of stars. Up to now, the most stringent constraints have been
provided by helioseismology and stellar cluster studies. However, the
contribution of asteroseismology becomes more and more important, giving
us an opportunity to probe the interiors and atmospheres of very different
stellar objects. A variety of pulsating variables allows us to test
various parameters of micro- and macrophysics by means of oscillation
data. I will review the most outstanding key problems, both observational
and theoretical, of which our knowledge can be improved by means of
asteroseismology.
\end{abstract}


\section*{Introduction}

Stars are the main components of the visible Universe. Our understanding
of their internal structure and the way they evolve is a crucial piece in
understanding how galaxies form and evolve. Studies of stellar clusters
bring general information about the composition and evolution of stars but
they are not sensitive enough to teach us about microphysical processes
determining stellar structure. The only particles that carry information
about the solar centre are neutrinos, which are a direct byproduct of
nuclear fusion reactions. As a detection of these weakly interacting
particles is very difficult, and for stars other than the Sun still beyond
our reach, pulsations provide the only opportunity for testing the physics 
of stellar interiors and, in a next step, theory of stellar evolution.

Helioseismology has by far the greatest contribution to theory of stellar
structure and evolution. The global studies led to determinations of the
solar age, the depth of the convective zone, helium abundance and
rotational profile. The great impact of helioseismology concerns also
atomic physics, exemplified by its role in solving the solar neutrino
problem, or by testing opacity data and equation of state. Now a new era
is opening up for local helioseismology, which should provide
three-dimensional maps of the solar interior and magnetic field.

Pulsating variables cover a wide range of masses and every stage in
stellar evolution. As a consequence we can observe pulsations with various
periods, amplitudes and shapes of light curves which result from
excitation of different modes. In spite of this variety, there are only
two underlying mechanisms for driving stellar pulsations. The first is
self excitation in the layers which operate as a heat engine. This
instability mechanism excites pulsations in most stars, beginning from
classical instability strip stars, through B~type main sequence stars, hot
subdwarfs to white dwarfs. The second way to make a star pulsate is to
force stochastic oscillations by turbulent convection. This stochastic
excitation drives solar-like oscillations, including those observed in the
Sun, and is expected in all stars with extended convective outer layers.
The most complete version of the Hertzsprung-Russell diagram of pulsating
stars is shown in Fig.~1. This diagram was constructed by Simon Jeffery
on the basis of an idea of J{\o}rgen Christensen-Dalsgaard.

The diversity of stellar pulsations offers an opportunity to probe various
physical phenomena, like element mixing, opacity, efficiency of
convection, magnetic field and non-uniform rotation. The ultimate goal of
asteroseismology is to construct a seismic model which reproduces all
observed frequencies and the corresponding pulsational mode
characteristics. During the last few years we are witnessing a growing
impact of asteroseismology in extracting constraints on stellar parameters
and physical processes in the stellar interiors. Another door to the
application of asteroseismology is opening by pulsating stars which harbour
planets, like $\mu$ Ara, a solar-like star (Bazot et al.\ 2005) or V391
Peg, a hot subdwarf (Silvotti et al.\ 2007).

In the first section, I will list observational key problems, which I
consider most challenging for theory of stellar evolution and pulsation.
The second section is devoted to the most puzzling theoretical aspects.
Conclusions end this review. Because of the space limit, I had to make a
crude selection. Therefore, I did not discuss, for example, the problems
connected with the presence of magnetic field and the potential of
asteroseismology to test it. These issues can be found, e.g., in a review
by Kochukhov of the roAp pulsators (these proceedings).
	
\figureCoAst{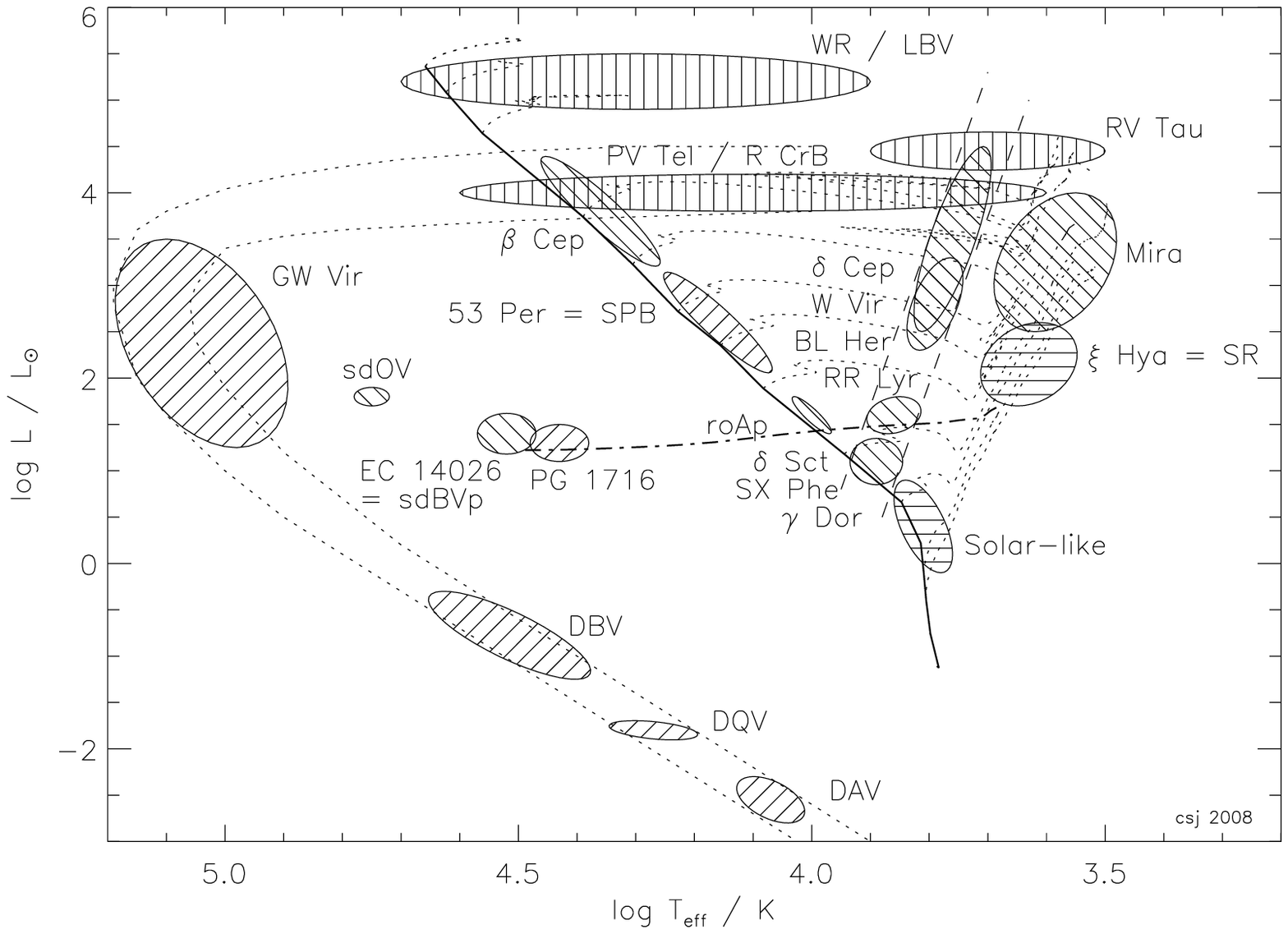}{The Hertzsprung-Russell diagram with
schematic locations of various types of pulsating variables. The hatching
styles correspond to different types of stellar pulsation: diagonal lines
$-$ heat-driven p and g~mode pulsations, horizontal lines $-$ solar-like
oscillations, and vertical lines $-$ strange mode (highly non-adiabatic)
pulsations. The zero age main sequence is depicted by a heavy solid line
and the zero age horizontal branch by a heavy dot-dash line. Evolutionary
tracks and white dwarf cooling sequences are shown as dotted lines. Figure
courtesy of Simon Jeffery (2008a).}
{puls_jadwiga}{ht!}{clip,angle=0,width=\textwidth}

\section*{Observational challenges}

\subsection*{Classical Cepheids}

Classical Cepheids play a crucial role in astronomy and their importance
has been recognized a long time ago. Firstly, they are primary distance
indicators on the extragalactic scale thanks to the Period Luminosity
Relation discovered by Henrietta Leavitt in 1908. Secondly, their
pulsations provide a test of stellar evolution theory of intermediate mass
stars.

The longest standing problem connected with these variables was a
discrepancy for double mode Cepheids between masses estimated from
pulsation and evolution theory. The implementation of new opacity data
(Iglesias \& Rogers 1991) made by Moskalik et al.\ (1992) allowed
essentially to reconcile this mass discrepancy. Despite this crucial work
some disagreement still remains and possible sources of this persistent
problem are opacities, mass loss and internal mixing processes (Keller
2008 and reference therein). Recently, yet another problem arose for
double mode Cepheid models. Smolec \& Moskalik (2008a, b) have shown that
if buoyancy forces in convectively stable layers are included in pulsation
modelling, double mode solutions do not exist.

During the last few years a number of interesting and challenging discoveries
were made, based mostly on of the OGLE project data. There are:\\
$\bullet$ Blazhko Cepheids (Moskalik \& Ko{\l}aczkowski 2008a)\\
$\bullet$ non-radial modes in classical Cepheids (Moskalik \& Ko{\l}aczkowski 2008b)\\
$\bullet$ 1O/3O double-mode Cepheids (Soszy\'nski et al.\ 2008a)\\
$\bullet$ single-mode second-overtone Cepheids (Udalski et al.\ 1999, Soszy\'nski et al.\ 2008b),\\
$\bullet$ triple-mode Cepheids (Moskalik et al.\ 2004, Soszy\'nski et al.\ 2008a, b)\\
$\bullet$ eclipsing binary systems containing Cepheids (Soszy\'nski et al.\ 2008b).

All the above theoretical and observational facts wait for explanation and
call for new generation models of stellar pulsation and evolution.

\subsection*{B type main sequence stars}

Studying B type main sequence stars is of great importance because those
with masses greater than about 8\,M$_\odot$ are progenitors of Type II
Supernovae, whereas those with masses less than about 8\,M$_\odot$ form
the CNO elements in the Universe. In some of these objects pulsation
occurs, giving an opportunity to study physics of stellar interiors by
means of oscillation data. There are two classes of B type main sequence
pulsators: 1) $\beta$ Cephei stars with masses larger than 8\,M$_\odot$
and spectral types B0$-$B2.5, in which mainly pressure (p) modes are
excited, and 2) Slowly Pulsating B type (SPB) stars with masses smaller
than 8\,M$_\odot$ and spectral types B3$-$B9, which pulsate in high order
gravity (g) modes.

Many years had elapsed since the discovery of $\beta$ Cephei stars before
the cause of their pulsations was identified. As in the case of the
classical Cepheid mass discrepancy problem, the reason was that the Los
Alamos opacities missed the metal-bump because of underestimations of the
heavy element contributions, mostly from the iron group (e.g.\ Moskalik \&
Dziembowski 1992, Dziembowski et al.\ 1993). As pulsations in $\beta$
Cephei and SPB stars are strictly connected with the metal opacity bump,
nobody expected to find many of them in the Magellanic Clouds (MCs),
because the metallicity in the MCs is much lower than in the Milky Way;  
$Z=0.007$ for the Large Magellanic Cloud and $Z=0.002$ for the Small
Magellanic Cloud. Therefore, the discovery of a large number of $\beta$
Cep and SPB variables in the MCs was rather a surprise. This started with
the pioneering paper by Pigulski and Ko{\l}aczkowski (2002). The later
papers by Ko{\l}aczkowski et al.\ (2006) and Karoff et al.\ (2008)
increased the number of known B type main sequence pulsators in the MCs.
Moreover, the newly determined solar chemical composition (Asplund
et al.\ 2005, AGS05) gives a
lower metallicity of about $Z=0.012$ compared to the older one 
(Grevesse \& Noels 1993) 
$Z=0.017$. Nevertheless, the new $Z$ value does not cause disappearance of
the B main sequence instability strip, as one would initially expect,
because most of the reduction was obtained in the CNO elements and now the
relative abundances of Fe group elements are significantly higher than the
older ones (Pamyatnykh \& Ziomek 2007, Miglio et al.\ 2007). However, the
problem with the existence of $\beta$ Cep stars in SMC remains open. This
observational fact waits for an explanation and indicates that
improvements are still needed in the treatment of opacities, mixing
processes, diffusion etc., as well as more observational data should be
gathered.

Another challenge is the occurrence of hybrid B type pulsators like $\nu$
Eri and 12 Lac (e.g., see Handler, these proceedings). The problem is that
the lower frequencies observed in the g~mode range cannot reach the
instability in pulsational models. This instability problem occurs also
for very high frequency modes. These two stars were recently studied by
Dziembowski \& Pamyatnykh (2008), who highlighted problems of mode
excitation, uncertainties in opacity data and element distribution, extent
of the overshooting distance and internal rotation.

\subsection*{Hot subdwarfs}

B type subdwarfs are stars in the core helium burning phase with a thin
hydrogen envelope. No hydrogen shell burning takes place. These stars
represent the final stage before the white dwarf phase but only 2\% of
white dwarfs are formed through this channel.

Short period pulsations in these objects were theoretically predicted by
Charpinet et al.\ (1997); Kilkenny et al.\ (1997) found them
observationally. Then, Green et al.\ (2003) found B type subdwarfs
pulsating with long periods. Both types of pulsation are driven by the
$\kappa$~mechanism due to the Z-opacity bump. Fontaine et al.\ (2006)
incorporated a nonuniform iron profile as determined by the condition of
diffusive equilibrium between gravitational settling and radiative
levitation. This allowed removing the artificial assumption of a very high
value of the metallicity. The evolutionary status of subluminous B type
stars seems to be proven but not their origin. There are two most
supported scenarios: 1) single star evolution, or 2) binary star
evolution, which can be a common envelope evolution, a stable Roche lobe
overflow or a merger of two He WD stars.

Recently, also pulsation in O~type subdwarfs was discovered (Woudt et al.\
2006, Rodriguez-Lopez et al.\ 2007). The sdO stars are the more evolved and
even more puzzling cousins of sdB stars. These objects have a C/O core and
are in the helium burning shell phase. As they cover quite a large part of
the HR diagram, they are thought to have many different origins. The
,,luminous'' (low gravity) sdO stars are believed to be post-AGB stars.
For the ,,compact'' (high gravity) sdO stars two origin scenarios were
proposed depending on the helium abundance: 1) post~EHB objects
(descendants of sdBs) for He-deficient sdOs, and 2) a merger of two He WDs
or a delayed core He flash scenario for He-enriched sdOs (Rodriguez-Lopez
et al.\ 2007). Fontaine et al.\ (2008) performed non-adiabatic pulsational
calculations taking into account time-dependent diffusion, as has been
done before for sdB pulsators. They showed that radiative levitation
causes pulsational instabilities in the sdO stars through an iron
accumulation in the driving region.

There is great potential for asteroseismology of OB subdwarf pulsators
because their radiative atmospheres combined with high gravities make them
ideal for investigating diffusive processes. The problem with each origin
scenario of sdOB stars is that a large amount of hydrogen has to be lost
before or just at the beginning of helium core ignition. As in the case of
the main sequence B type pulsators, additional opportunities come from
hybrid sdOB pulsators (Schuh et al.\ 2006, Lutz et al.\ 2008).

\subsection*{Extreme helium stars}

The extreme helium stars are low mass highly evolved objects (supergiants)
of B and A types, with spectra showing very weak or no hydrogen lines
(Jeffery 2008b). They cover a wide range of effective temperature and most
of them have luminosities close to the Eddington limit. The fundamental
questions about this kind of stars are their origin (how to remove a
hydrogen-rich envelope) and their connection with stars with normal He
abundance. Two main origin scenarios were proposed. In the first scenario,
the EHe star is a direct product of a single white dwarf which underwent a
late thermal pulse (e.g.\ Iben et al.\ 1983). The second scenario involves
a white dwarf binary which has merged (e.g.\ Webbink 1984).

The existence of EHe stars exhibiting pulsational variability is extremely
significant. Jeffery (2008c) introduced consistent classification of
variable EHe stars. The first group consists of PV~Tel stars with strange
mode pulsational instability (Saio \& Jeffery 1988) which is supposed to
be present in all stars with sufficiently high luminosity/mass ratio. The
second class includes BX Cir stars with pulsations driven by the $\kappa$
mechanism operating in the Z~bump (Saio 1993). Here, the role of the iron
group element opacity is increased due to a reduction of hydrogen (Jeffery
\& Saio 1999).

There are also helium-rich subluminous OB stars whose origin is unclear
(Napiwotzki 2008, Ahmad \& Jeffery 2008), but the hypothesis of merging
two He~core white dwarfs seems to be preferred. Questions arise about
their connection with hydrogen-rich sdOB stars or with EHe stars. In the
light curve of one He~sdB object, Ahmad \& Jeffery (2005) detected
multiperiodic variations. These authors have shown that the variations
can be associated with high order g~mode pulsations, but according to
theory such modes are stable.

\subsection*{New class of white dwarfs}

Asteroseismology of white dwarf (WD) pulsators provides many constraints
on stellar physics and evolution. For example, from period changes we can
estimate the cooling rate, which in turn can give information on the age
of our Galaxy and can measure the neutrino generation rate in hot WDs
(Winget 1998). WD asteroseismology also supplies a unique test of the
equation of state of matter at high densities and temperatures. Because
most stars (97\%) will end up as WDs, their studies are of special
importance. An excellent review on the three types of WD pulsators, i.e.,
ZZ Cet (DAV), V777 Her (DBV) and GW Vir (PNNV+DOV) type stars, can be
found in Fontaine \& Brassard (2008).

WDs of DA spectral type have hydrogen-rich atmospheres. The atmospheres of
the two other types (DB, DO$+$PNNV) are dominated by helium. There is one
more class of WDs at $T_{\rm eff}~11 000-13 000$~K, with carbon lines in
their spectra, designated the DQ type, but until recently even in this
class, helium was believed to be the dominant component of the atmosphere.
Therefore, the discovery of hot DQ WDs ($T_{\rm eff}~18 000-23 000$~K)
with carbon-dominated atmospheres and little or no H and He, was rather
unexpected (Dufour et al.\ 2007). The origin of the cooler DQ WDs is quite
well understood in the framework of the model of carbon dredge-up by the
deep helium convection zone. On the other hand, the origin of hot DQ WDs
cannot be explained within any known post-AGB evolution channel.

Soon thereafter, Fontaine et al.\ (2008) have explored the instability of
hot DQ WD models against pulsation. Their analysis showed that gravity
modes should be excited in such models in the period range $100-700$\,s.
Almost at the same time, Montgomery et al.\ (2008) announced the
photometric variability of one carbon-atmosphere object with a period of
417\,s. This makes this object a prototype of a new class of pulsating
WDs. Recently, two more variable DQ WDs have been discovered by Barlow et
al.\ (2008). Having in mind the potential of asteroseismology, one can
expect interesting results from seismic studies of hot DQ WD pulsators
before long.

\section*{Theoretical challenges}

\subsection*{Opacities and chemical composition}

Stellar opacities constitute a vital component in modelling stellar
structure because they determine the transport of radiation through
matter. Their values depend on the temperature, density and chemical
composition, $\kappa(T,\rho, X_i)$. For a long time the only source of the
opacity data for astrophysical purposes was the Los Alamos Opacity Library
(LAOL). The LAOL was in use until 1990, although already in 1982 N.~R.\
Simon dared to blame them for the failure in explaining the Cepheid mass
discrepancy and pulsations of $\beta$ Cephei stars, and urged
reexamination of this crucial input (Simon 1982). Finally, in early
90ties, two teams of atomic physicists recomputed the opacity data and
found that the LAOL underestimated the contributions from the heavy
elements by a factor of $2-3$ at a temperature of about 200\,000\,K. The
first team was represented by Iglesias \& Rogers (e.g., 1991, 1996), who
called their results OPAL (OPAcity Library). The second was an
international team led by M.\ J.\ Seaton and their opacities were named OP
(Opacity Project) (e.g.\ Seaton 1992, 1996, 2007).

The computation of new opacity tables was a milestone for astrophysics and
the main consequences of a huge enhancement of the metal opacity bump
were: 1) the seismic model of the Sun was improved, 2) the Cepheid mass
discrepancy was significantly reduced, 3) the pulsations of B type main
sequence stars and of 4) some extreme He stars were explained, 5)
pulsations of sdB and sdO stars were predicted.

The opacity data are being constantly updated but the main features are
kept unchanged. Recently, a substantial revision of the solar chemical
mixture has led to a significant change of the opacity values (AGS05). 
The new solar composition was reduced
mostly in the CNO elements and now the solar metallicity is only 70\% of
the older one by Grevesse \& Noels (1993). As mentioned before, these new
solar abundances did not diminish the extent of the pulsational
instability strip of B type stars because the relative Fe~abundance was
increased, hence the role of the Z~bump opacity in driving pulsations was
amplified. There is other strong evidence supporting the AGS05 mixture.
For example, now the solar metallicity is in better agreement with
metallicities of stars in its neighbourhood. Then, galactic beat Cepheid
models with the AGS05 abundances better fit the observations (Buchler \&
Szabo 2007). Moreover, with reduced $Z$, the pre-main sequence models show
smaller lithium depletion (Montalb\'an \& D'Antona 2006) which brings them
closer to the observations. However, the new solar composition had bad
consequences for the helioseismic model which lost consistency with the
standard solar model. This problem was extensively discussed in a review
by Basu \& Antia (2008).

There are many papers in which stellar pulsations of various objects were
used as a test of opacity data, e.g., Dziembowski \& Pamyatnykh (2008) for
$\beta$ Cep stars, Jeffery \& Saio (2006) for pulsating subdwarf B stars,
Lenz et al.\ (2008) for $\delta$ Sct stars or Th\'eado et al.\ (2008) for roAp
stars. The role of input from opacity data in pulsation computations was
recently summed up by Montalb\'an \& Miglio (2008).

\subsection*{Rotation}

A credible theory of stellar structure and evolution should incorporate
rotation. Firstly, rotation affects the stellar structure by breaking
spherical symmetry. The most extreme example known is Achernar with a
ratio of the major to minor axes equal to $1.56 \pm 0.05$. Secondly,
rotation activates various processes, like meridional circulation, shear
instabilities, diffusion, horizontal turbulence, which cause mixing.
Thirdly, the distribution of internal angular momentum is determined
through different rotational velocity at different depths. And finally,
mass loss from the surface can be enhanced by rapid rotation through
centrifugal effects. Many excellent papers on stellar rotation have been
published by, e.g., Sweet, \"Opik, Tassoul, Roxburgh, Zahn, Spruit,
Deupree, Talon, Meynet, Maeder, Mathis, but there is no space here to
mention the results of all of them. In the last years the most
comprehensive studies of the rotational effects on stellar structure and
evolution were presented in a series of papers by Maeder, Meynet and their
collaborators (e.g., Maeder \& Meynet 2000a, b).

How rotation affects pulsation depends on the rotation rate and on the
closeness of oscillation frequencies. If the pulsational frequency is much
larger than the rotational angular velocity, the perturbation approach is
applicable; this considerably simplifies calculations and in the first
order approximation each pulsational mode can be described by a single
spherical harmonic. This is no longer true if the higher order effects of
rotation are included (Dziembowski \& Goode 1992). Moderate rotation can
also couple modes for which the frequency distance is close to the
rotational angular velocity and where the spherical harmonic indices
satisfy the relations: $l_j = l_k+2$ and $m_j = m_k$ (Soufi et al.\
1998). The third order expression for a rotationally split frequency can
by found in Goupil et al.\ (2000). As for mode geometry, the main effect of
fast rotation is the confinement of pulsation towards the stellar equator
(e.g.\ Townsend 1997).

An entirely different approach has to be applied if the pulsation
frequencies are of the order of or smaller than the rotation frequency
($\nu_{\rm puls} \sim \nu_{\rm rot}$). Three different treatments of slow
modes can be found in the literature: the traditional approximation
(Townsend 2003), expansion in Legendre function series (Lee \& Saio 1997)
and 2D$(r,\theta)$ modelling (Savonije 2007).

Rotation also complicates the identification of pulsation modes from
photometric diagnostic diagrams because they become dependent on
inclination angle, $i$, azimuthal order, $m$, and rotation velocity,
$v_{\rm rot}$. This problem was studied by Daszy\'nska-Daszkiewicz et al.\
(2002) for coupled modes, and by Townsend (2003) and
Daszy\'nska-Daszkiewicz et al.\ (2007) for slow modes.

The asteroseismic potential of rotating pulsators lies in the rotational
splitting kernel, $K(r)$ that gives information on the rotational
profile $\Omega(r)$. The result that the rotation rate increases inward
was obtained from studies of many pulsating stars, e.g., Goupil et al.\
(1993) for the $\delta$~Sct star GX~Peg; Dziembowski \& Jerzykiewicz
(1996) for the $\beta$~Cep star 16~Lac; Aerts et al.\ (2003) for the
$\beta$~Cep star V836~Cen, Pamyatnykh et al.\ (2004) for the $\beta$~Cep
star $\nu$~Eri and Dziembowski \& Pamyatnykh (2008) for the $\beta$~Cep
stars $\nu$~Eri and 12~Lac.

Another question is about the impact of pulsation on rotational evolution.
Talon \& Charbonnel (2005) showed that internal gravity waves contribute
to braking rotation in the inner regions of low mass stars. Computations
by Townsend \& MacDonald (2008) demonstrated that pulsation modes can
redistribute angular momentum and trigger shear instability mixing in the
$\mu$~gradient zone. Mathis et al.\ (2008) discussed the transport of
angular momentum in the solar radiative zone by internal gravity waves.

\subsection*{Convection}

Convection plays a very important role in the transport of energy and
mixing of matter. It is a complex physical process with a
three-dimensional, non-local and time-dependent character. The convective
cells are also a source of acoustic waves in subphotospheric layers,
leading to stochastic excitation of stellar oscillations. In turn,
dissipation of acoustic energy heats stellar chromospheres, causing
"fingerprints" in spectral lines. Also stellar activity is a result of
joint action of convection and differential rotation.

The most widely used description of stellar convection is the Mixing
Length Theory (MLT) due to B\"ohm-Vitense (1958) and some modifications of
it. In the framework of MLT, the size of convective elements is
parameterized by the the mixing length parameter, $\alpha_{\rm conv}$,
which is adjusted to fit some observational quantities. A step forward was
done by Canuto et al.\ (1996), who formulated a theory of turbulent
convection taking into account the full spectrum of convective eddies. A
3D hydrodynamical simulation by Stein \& Nordlund (1998) allowed to
reproduce qualitatively convection in the solar surface layers.

A discussion of differences between the 3D hydrodynamical model of
convection and 1D MLT model was presented by Steffen (2007). The main
results of his comparisons were that it is impossible to reproduce the
correct temperature profile with any value of the MLT parameter, and that
the radiative layer between two convection zones is completely mixed.

The presence of pulsation in stars with expanded convective envelopes
complicates the picture even more. This is the case for Classical
Cepheids, RR Lyrae stars, red giant pulsators, $\delta$ Scuti and $\gamma$
Doradus stars, as well as for pulsating white dwarfs of the V777~Her and
ZZ~Cet types. A simplistic approach to pulsational modelling of these
objects is a convective flux freezing approximation which assumes that the
convective flux is constant during the pulsation cycle. Although in some
$\delta$ Scuti stars convection seems to be inefficient
(Daszy\'nska-Daszkiewicz et al.\ 2003, 2005a), this is a crude
approximation and an adequate treatment of pulsation-convection
interaction should be applied.

The first formulation of pulsation-convection interactions in which
convection is non-local and time-dependent was given by Unno (1967) and
Gough (1977). Following this concept and its variants, many research
groups undertook attempts to model stellar oscillations in various
objects: solar-like stars (Houdek, Goupil, Samadi), $\delta$ Scuti and
$\gamma$~Doradus stars (Xiong, Houdek, Dupret, Grigahcene, Moya),
Classical Cepheids and RR Lyr variables (Feuchtinger, Stellingwerf,
Buchler, Kollath, Smolec, Moskalik), pulsating red giants (Xiong, Deng,
Cheng) and recently, V777 Her white dwarfs (DBV) (Quirion, Dupret).

The potential of asteroseismology for estimating the size of the
convective core in massive stars was also explored. Dziembowski \&
Pamyatnykh (1991) demonstrated that modes which are largely trapped in the
region surrounding the convective core boundary can measure the extent of
overshooting. The first evidence of core overshooting in a $\beta$ Cep
star was found by Aerts et al.\ (2003) for V836 Cen. Miglio et al.\ (2008)
studied the sensitivity of high order g~modes in SPB and $\gamma$ Dor
stars to the properties of convective cores. In particular, the period
spacing of gravity modes depends on the location and shape of the chemical
composition gradient.

In general, convective transport of energy above the core in OB type stars
is negligible. However, recently Maeder, Georgy \& Meynet (2008) have
shown that very fast rotation in massive OB stars can increase the extent
of outer convective envelopes. This has many consequences, e.g., acoustic
modes can be generated. Also, stars close to the Eddington limit may
develop a convective envelope.

\subsection*{Mass loss}

Mass loss occurs in all late evolutionary phases and in massive stars.
There are two principal mechanisms driving stellar winds: radiation
driving in hot stars and dust driving in cool and luminous stars (e.g.\
Owocki 2004). For most stars, no consistent description of the mass loss
rate exists and empirical formulae are used instead. Also, the interaction
between mass loss and rotation has not yet been fully understood (e.g.\
Maeder \& Meynet 2004, Owocki 2008). Some stars with strong winds exhibit
also pulsations. These are Mira and semi-regular (SR) variables, massive
OB main sequence stars, Wolf-Rayet stars and Luminous Blue Variables
(LBV). For such objects a natural question emerges about pulsation and
mass loss coupling. It has been recognized many years ago that pulsations
in large amplitude variables, like Miras or Cepheids, can enhance mass
loss (e.g.\ Castro 1981, Wood 2007, Neilson \& Lester 2008). Constraints
have been derived from relations between the mass loss rate and
pulsational period and between the wind outflow speed and pulsational
period (e.g., Knapp et al.\ 1998). As for hot pulsators, Howarth et al.\
(1993) found wind variability in $\zeta$ Oph with the pulsational cycle,
and Kaufer (2006) detected the pulsation beat period in H$\alpha$ profile
observations for a B0~type supergiant. A nice discussion of the coupling
between mass loss and pulsation in massive stars can be found in Townsend
(2007).

Interesting results have been obtained recently by Quirion, Fontaine \&
Brassard (2007) for hot white dwarf pulsators of the GW Vir type. The
authors showed that useful constraints on mass loss can be inferred from
the red~edge position of the instability strip.

\section*{Conclusions}

An ultimate goal of asteroseismology is to help solving the equation
$observation=theory$ and to avoid the equation
$more~data=less~understanding$. But as we are all aware, a more realistic
treatment of macro- and microphysics in stellar modelling would be
desirable.

One should have in mind that not only pulsation frequencies can probe
stellar structure. An ideal seismic stellar model should account both for
all measured oscillation frequencies and for associated pulsation mode
characteristics. Therefore more use should be made of photometric and
spectroscopic observables, i.e., amplitudes and phases of photometric and
spectroscopic variations, and simultaneous multi-colour photometric and
spectroscopic observations should be carried out. From such data we can
infer an additional asteroseismic constraint, which is the ratio of the
bolometric flux variations to the radial displacement. This new
asteroseismic probe, called the $f$ parameter, is determined in
subphotospheric layers, and therefore it is complementary to the frequency
data which poorly probe these stellar regions. The value of $f$ is very
sensitive to global stellar parameters, element mixture (hence the mixing
processes), opacities and subphotospheric convection. Therefore, a
comparison of empirical and theoretical values of $f$ provides additional
stringent constraints on various physical parameters and processes. Such
asteroseismic studies were proposed by Daszynska-Daszkiewicz et al.\ (2003,
2005ab) and successfully applied to $\delta$ Scuti and $\beta$ Cephei
stars. In the case of $\delta$ Scuti stars, useful constraints on
subphotospheric convection were derived, and in the case of $\beta$ Cep
stars, on opacities.

In the next step, the $f$ parameter should be fitted to observations
together with the pulsation frequency. These complex asteroseismic studies
should lead to stronger constraints, improving our knowledge in theory of
stellar pulsation and evolution.

\acknowledgments{The author would like to thank the SOC members of the
JENAM 2008 Symposium No.~4 for their invitation and Miko{\l}aj
Jerzykiewicz and Alosza Pamyatnykh for carefully reading the manuscript
and their comments. The EC is acknowledged for the establishment of the
European Helio- and Asteroseismology Network (HELAS, No.~026138), which
made the authors' participation at this meeting possible.}

\References{
\rfr Aerts, C., Thoul, A., Daszy\'nska, J., et~al.\ 2003, Science, 300, 1926
\rfr Ahmad, A., \& Jeffery, C.~S.\ 2005, A\&A, 437, L51
\rfr Ahmad, A., \& Jeffery, C.~S.\ 2008, 
     in ''Hydrogen-Deficient Stars'', eds.\ K.\ Werner and T.\ Rauch,
     ASP Conf.\ Ser., 391, 261
\rfr Asplund, M., Grevesse, N., Sauval, A.~J., et~al.\ 2005, A\&A, 431, 693
\rfr Barlow, B.~N., Dunlap, B.~H., Rosen, R., \& Clemens, J.~C.\ 2008, ApJ, 2008, 688, L95
\rfr Basu, S., \& Antia, H.~M.\ 2008, Phys.\ Rep., 457, 217
\rfr Bazot, M., Vauclair, S., Bouchy, F., \& Santos, N.~C.\ 2005, A\&A, 440, 615
\rfr B\"ohm-Vitense, E.\ 1958, Z.\ Astrophys., 46, 108
\rfr Buchler, J.~R., \& Szabo, R.\ 2007, ApJ, 660, 723
\rfr Canuto, V.~M., Goldman, I., \& Mazzitelli, I.\ 1996, ApJ, 473, 550
\rfr Charpinet, S., Fontaine, G., Brassard, P., et~al.\ 1997, ApJ, 483, L123
\rfr Daszy\'nska-Daszkiewicz, J., Dziembowski, W.~A., Pamyatnykh, A.~A., 
\& Goupil, M.-J.\ 2002, A\&A, 392, 151
\rfr Daszy\'nska-Daszkiewicz, J., Dziembowski, W.~A., \& Pamyatnykh, A.~A.\ 
2003, A\&A, 407, 999
\rfr Daszy\'nska-Daszkiewicz, J., Dziembowski, W.~A., Pamyatnykh, A.~A., 
et~al.\ 2005a, A\&A, 438, 653
\rfr Daszy\'nska-Daszkiewicz, J., Dziembowski, W.~A., \& Pamyatnykh, A.~A.\ 
2005b, A\&A, 441, 641
\rfr Daszy\'nska-Daszkiewicz, J., Dziembowski, W.~A., \& Pamyatnykh, A.~A.\ 
2007, AcA, 57, 11
\rfr Dufour, P., Liebert, J., Fontaine, G., \& Behara, N.\ 2007, Nature 450, 
522
\rfr Dziembowski, W.~A., \& Goode, P.~R.\ 1992, ApJ, 394, 670
\rfr Dziembowski, W.~A., \& Jerzykiewicz, M.\ 1996, A\&A, 306, 436
\rfr Dziembowski, W.~A., \& Pamyatnykh, A.~A.\ 1991, A\&A 248, L11
\rfr Dziembowski, W.~A., \& Pamyatnykh, A.~A.\ 1993, MNRAS, 226, 204
\rfr Dziembowski, W.~A., \& Pamyatnykh, A.~A.\ 2008, MNRAS, 385, 2061
\rfr Fontaine, G., Brassard, P., Charpinet, S., \& Chayer, P.\ 2006, MmSAI, 77, 49
\rfr Fontaine, G., \& Brassard, P.\ 2008, PASP, 120, 104
\rfr Fontaine, G., Brassard, P., Green, B., et~al.\ 2008, A\&A, 486, L39
\rfr Fontaine, G., Brassard, P., \& Dufour, P.\ 2008, A\&A, 483, L1
\rfr Goupil, M.-J., Dziembowski, W.~A., Pamyatnykh, A.~A., \& Talon, S.\ 2000, 
     in ''Delta Scuti and Related Stars'', 
     eds.\ M.\ Breger and M.~H.\ Montgomery, 
     ASP Conf.\ Ser., 210, 267
\rfr Grevesse, N., Noels, A.\ 1993, 
     in ''Origin and Evolution of the Elements'', 
     eds.\ N.~Pratzo, E.~Vangioni-Flam, and M.~Casse, 
     Cambridge Univ.\ Press, Cambridge, p.~15
\rfr Howarth I.~D., Bolton, C.~T., Crowe, R.~A., et~al.\ 1993, ApJ, 417, 338
\rfr Iben, J., Kaler, J.~B., Truran, J.~W., \& Renzini, A.\ 1983, ApJ, 264, 605
\rfr Iglesias, C.~A., \& Rogers, F.~J.\ 1991, ApJ, 371, 408
\rfr Iglesias, C.~A., \& Rogers, F.~J.\ 1996, ApJ, 464, 943
\rfr Jeffery, C.~S., Drilling, J.~S., Harrison, P.~M., et~al.\ 1997, A\&AS, 125, 501
\rfr Jeffery, C.~S., \& Saio, H.\ 1999, MNRAS,308, 221
\rfr Jeffery, C.~S., \& Saio, H.\ 2006, MNRAS,372, L48
\rfr Jeffery, C.~S.\ 2008a, CoAst, 157, 240
\rfr Jeffery, C.~S.\ 2008b, 
     in ''Hydrogen-Deficient Stars'', eds.\ K.\ Werner and T.\ Rauch, 
     ASP Conf.\ Ser., 391, 53 
\rfr Jeffery, C.~S.\ 2008c, IBVS, 5817, 1
\rfr Karoff, C., Arentoft, T., Glowienka, L., et~al.\ 2008, MNRAS, 386, 1085
\rfr Keller, S.~C.\ 2008, ApJ, 677, 483
\rfr Kochukhov, O.\ 2009, CoAst, 159, \pageref{ref:kochukhov}
\rfr Ko{\l}aczkowski, Z., Pigulski, A., Soszy\'nski, I., et~al.\ 2006, 
MmSAI, 77, 336
\rfr Kilkenny, D., Koen, C., O'Donoghue, D., \& Stobie, R.~S.\ 1997, MNRAS, 
285, 640
\rfr Knapp, G.~R., Young, K., Lee, E., \& Jorissen, A.\ 1998, ApJS, 117, 209
\rfr Lenz, P., Pamyatnykh, A.~A., Breger, M., \& Antoci, V.\ 2008, A\&A, 478, 
855
\rfr Lutz, R., Schuh, S., Silvotti, R., et~al.\ 2008, 
     in ''Hot Subdwarf Stars and Related Objects'',
     eds.\ U.~Heber, C.~S.\ Jeffery, and R.\ Napiwotzki,
     ASP Conf.\ Ser., 392, 339
\rfr Maeder, A., Georgy, C., \& Meynet, G.\ 2008, A\&A, 479, L37
\rfr Mathis, S., Talon, S., Pantillon, F.-P., \& Zahn, J.-P.\ 2008, SoPh, 251, 101
\rfr Meynet, G., \& Maeder, A.\ 2000, A\&A, 361, 101
\rfr Maeder, A., \& Meynet, G.\ 2000, ARA\&A, 38, 143
\rfr Miglio, A., Montalb\'an, J., \& Dupret, M.-A.\ 2007, MNRAS, 375, L21
\rfr Montalb\'an, J., D'Antona, F.\ 2006, MNRAS, 370, 1823
\rfr Montalb\'an J., \& Miglio, A.\ 2008, CoAst, 157, 160
\rfr Montgomery, M.~H., Williams, A.~K., Winget, D.~E., et~al.\ 2008, ApJ, 678, L51
\rfr Moskalik, P., Buchler, J.~R., \& Marom, A.\ 1992, ApJ, 385, 685
\rfr Moskalik, P., \& Dziembowski, W.~A.\ 1992, A\&A, 256, L5
\rfr Moskalik, P., \& Ko{\l}aczkowski, Z.\ 2008a, arXiv:0807.0615
\rfr Moskalik, P., \& Ko{\l}aczkowski, Z.\ 2008b, arXiv:0807.0623
\rfr Moskalik, P., Ko{\l}aczkowski, Z., \& Mizerski, T.\ 2004, ASP Conf.\ Ser., 310, 498
\rfr Napiwotzki, R.\ 2008, 
     in ''Hydrogen-Deficient Stars'', eds.\ K.\ Werner and T.\ Rauch, 
     ASP Conf.\ Ser., 391, 257 
\rfr Neilson, H.~R., \& Lester, J.~B.\ 2008, ApJ, 684, 569
\rfr Owocki, S.\ 2004, 
     in ''Evolution of Massive Stars, Mass Loss and Winds'', 
     eds.\ M.~Heydari-Malayeri, Ph.~Stee, and J.-P.\ Zahn, 
     EAS Publ.\ Ser., 13, 163
\rfr Owocki, S.\ 2008, 
     in ''Mass Loss from Stars and the Evolution of Stellar Clusters'',
     eds.\ A.\ de Koter, L.~J.~Smith, and L.~B.~F.~M.\ Waters, 
     ASP Conf.\ Ser., 388, 57
\rfr Pamyatnykh, A.~A., \& Ziomek, W.\ 2007, CoAst, 150, 207
\rfr Pamyatnykh, A.~A., Handler, G., \& Dziembowski, W.~A.\ 2004, MNRAS, 350, 1022
\rfr Pigulski, A., \& Ko{\l}aczkowski, Z.\ 2002, A\&A, 388, 88
\rfr Quirion, P.~O., Fontaine, G., \& Brassard, P.\ 2007, CoAst, 150, 247
\rfr Rodriguez-Lopez, C., Ulla, A., \& Garrido, R.\ 2007, MNRAS, 379, 1123
\rfr Saio, H.\ 1993, MNRAS, 260, 465
\rfr Saio, H., \& Jeffery, C.~S.\ 1988, ApJ, 328, 714
\rfr Lee, U., \& Saio, H.\ 1997, ApJ., 491, 839
\rfr Savonije, G.~J.\ 2007, A\&A, 469, 1057
\rfr Seaton, M.~J.\ 1992, RMxAA, 23, 180
\rfr Seaton, M.~J.\ 1996, MNRAS, 279, 95
\rfr Seaton, M.~J.\ 2007, MNRAS, 382, 245
\rfr Schuh, S., Huber, J., Dreizler, S., et~al.\ 2006, A\&A, 445, L31
\rfr Silvotti, R., Schuh, S., Janulis, R., et~al.\ 2007, Nature 449, 189
\rfr Simon, R.~N.\ 1982, ApJ, 260, L87
\rfr Smolec, R., \& Moskalik, P.\ 2008a, AcA, 58, 233
\rfr Smolec, R., \& Moskalik, P.\ 2008b, AcA, 58, 193
\rfr Soufi, F., Goupil, M-J., \& Dziembowski, W.~A.\ 1998, A\&A, 334, 911
\rfr Soszy\'nski, I., Poleski, R., Udalski, A., et~al.\ 2008a, AcA, 58, 153
\rfr Soszy\'nski, I., Poleski, R., Udalski, A., et~al.\ 2008a, AcA, 58, 153
\rfr Stein, R.~F., \& Nordlund, A.\ 1998, ApJ, 499, 914
\rfr Steffen, M.\ 2007, 
     in ''Convection in Astrophysics'', 
     eds.\ F.\ Kupka, I.\ Roxburgh, and K.\ Chan, 
     IAU Symposium, 239, 36
\rfr Talon, S., \& Charbonnel, C.\ 2005, A\&A, 440, 981
\rfr Th\'eado, S., Dupret, M.-A., Noels, A., \& Ferguson, J.~W.\ 2008, A\&A, in press
\rfr Townsend, R.\ 1997, MNRAS, 284, 839
\rfr Townsend, R.\ 2003, MNRAS, 343, 125
\rfr Townsend, R.\ 2007, 
     in ''Unsolved Problems in Stellar Physics'', 
     AIP Conf.\ Proceedings, 948, 345 
\rfr Townsend, R., \& MacDonald, J.\ 2008, 
     in ''Massive Stars as Cosmic Engines'', 
     eds.\ F.~Bresolin, P.~A.~Crowther, and J.~Puls, 
     IAU Symposium, 250, 161
\rfr Udalski, A., Soszy\'nski, I., Szyma\'nski, M., et~al.\ 1999, AcA, 49, 45
\rfr Webbink, R.~F.\ 1984, ApJ, 277, 355
\rfr Winget, D.~E.\ 1998, J.\ Phys., Condensed Matter, 10, 11247
\rfr Wood, P.~R.\ 2007, 
     in ''From Stars to Galaxies: Building the Pieces to Build Up the Universe'',
     eds.\ A.~Vallenari, R.\ Tantalo, L.\ Portinari, and A.\ Moretti, 
     ASP Conf.\ Ser., 374, 47
\rfr Woudt, P.~A., Kilkenny, D., Zietsman, E., et~al.\ 2006, MNRAS, 371, 1397
}

\end{document}